\documentclass[journal]{IEEEtran}

\usepackage{color}
\usepackage[pdftex]{graphicx}
\usepackage{cite}
\usepackage{multirow}
\usepackage{graphicx}
\usepackage{booktabs}
\usepackage{mathtools}
\usepackage[none]{hyphenat}
\usepackage{url}
\usepackage{subcaption}

\usepackage{lscape}
\graphicspath{{images/}}

\begin{document}

\title{Challenges and Opportunities for Simultaneous Multi-functional Networks in the UHF Bands}

\author{Xavier~Vilajosana, Guillem Boquet, Joan Melià, Pere Tuset-Peiró, Borja Martinez, Ferran Adelantado
\thanks{This project has been co-financed by the Spanish Ministry of Science, Innovation and Universities through the project RF-VOLUTION (PID2021-122247OB-I00).}
\thanks{The authors are with the Wireless Networks (WiNe) Research Lab, Internet Interdisciplinary Institute (IN3), Universitat Oberta de Catalunya (UOC).} }

\markboth{}{}

\maketitle

\begin{abstract}
%%%
%\gb{ DEADLINE : 15th OF JULY }

%\gb{ Total words: 3700 (28-june 16:00) of 4500 max.}

%\gb{ Max. 6 combined figures and tables.}

%\gb{ Max. 15 references.}
%%%

Multi-functional wireless networks are rapidly evolving and aspire to become a promising attribute of the upcoming 6G networks. Enabling multiple simultaneous networking functions with a single radio fosters the development of more integrated and simpler equipment, overcoming design and technology barriers inherited from radio systems of the past.
We are seeing numerous trends exploiting these features in newly designed radios, such as those operating on the mmWave band. 
In this article, however, we carefully analyze the challenges and opportunities for multi-functional wireless networks in UHF bands, advocating the reuse of existing infrastructures and technologies, and exploring the possibilities of expanding their functionality without requiring architectural changes.
We believe that both modern and legacy technologies can be turned into multi-functional systems if the right scientific and technological challenges are properly addressed. This transformation can foster the development of new applications and extend the useful life of these systems, contributing to a more sustainable digitization by delaying equipment obsolescence.
\end{abstract}


\IEEEpeerreviewmaketitle

\section{Introduction}
\label{sec:introduction}
Multi-functional wireless networks based on radio frequency (RF) convergence (RFC) are gaining momentum thanks to the miniaturization of microwave electronics, increased capabilities to integrate radio electronic systems in physical and mechanical substrates, augmented in-node signal processing options, novel virtualized antenna designs, and extended edge computing capabilities that enable smart post-processing of signals~\cite{7782415}.
Research in radio systems operating in the millimeter wave bands, as well as similar emerging concepts in the terahertz bands, are opening up new sensing and communication opportunities based on ideas derived from the radar and image processing fields~\cite{9354629}.
The enormous bandwidth available in these bands, as well as the well-established knowledge in the aforementioned fields, are showing exciting opportunities to use radio signals as sensors, typically for presence, tracking, or detection~\cite{9354629}. 

Nowadays, a significant portion of the telecommunication systems operates within the ultra-high frequency (UHF) band thanks to the good trade-off between propagation characteristics and available bandwidth. Besides radio and television broadcasting, the most common technologies are the cellular networks (3/4/5G), global navigation satellite systems (GNSS), radio frequency identification systems (RFID), low power wide area networks (LPWANs) and wireless local and personal area networks (WLANs and WPANs), such as IEEE~802.11 (Wi-Fi), IEEE 802.15.1 (Bluetooth) and IEEE 802.15.4 (ZigBee), among others. Their widespread use and state of deployment demands to explore the possibilities of expanding their functionality analogously to what is being done in higher bands, without implying the replacement or implementation of new equipment. 
The dual use of communication infrastructures provided by multi-functional networks, will have a positive impact on the digitalization of industries and the automation of buildings and infrastructures; that is, verticals where infrastructure already exists, but new requirements must be met in addition to communication, such as positioning and sensing.
Consequently, it will improve the efficiency of systems while delaying their obsolescence, thus improving the economical aspects of technology and contributing to a more sustainable digitization process~\cite{cristina}. %citar paper cristina si esta al arxiv.

Nonetheless, due to the intrinsic physical nature of UHF signals, with decimeter-scale wavelengths and limited available bandwidth, outdoor and indoor propagation, as well as the massive use of some sub-bands, there are many design and implementation challenges that must be faced for a realistic applicability of the concept. For example, the function simultaneity, positioning accuracy, communication performance, battery life, etc. In this article, we review the possibilities for multi-functionality of current technologies operating in the UHF band and present the opportunities and challenges that remain in moving towards the full RFC concept. 
Specifically, we outline and discuss the possibilities of materializing the concept within the band, drawing some directions for next-generation RF systems. We then devise research directions towards the reuse and the augmentation of existing systems, while identifying research challenges and implementation barriers that need to be addressed to ensure that augmented systems can be built on legacy infrastructures.

The rest of the article is organized as follows:
Section~\ref{sec:multifuncradiosys} describes the building blocks and core functions of a multi-functional radio system, matching it to the common technologies in the band.
Section~\ref{sec:applications} presents the different applications of RFC.
Section~\ref{sec:challenges} analyzes their challenges, limitations, and opportunities.
Finally, Section~\ref{sec:conclusions} concludes the article, providing the final remarks. 

\begin{figure*}[]
    \centering
    \includegraphics[width=\linewidth]{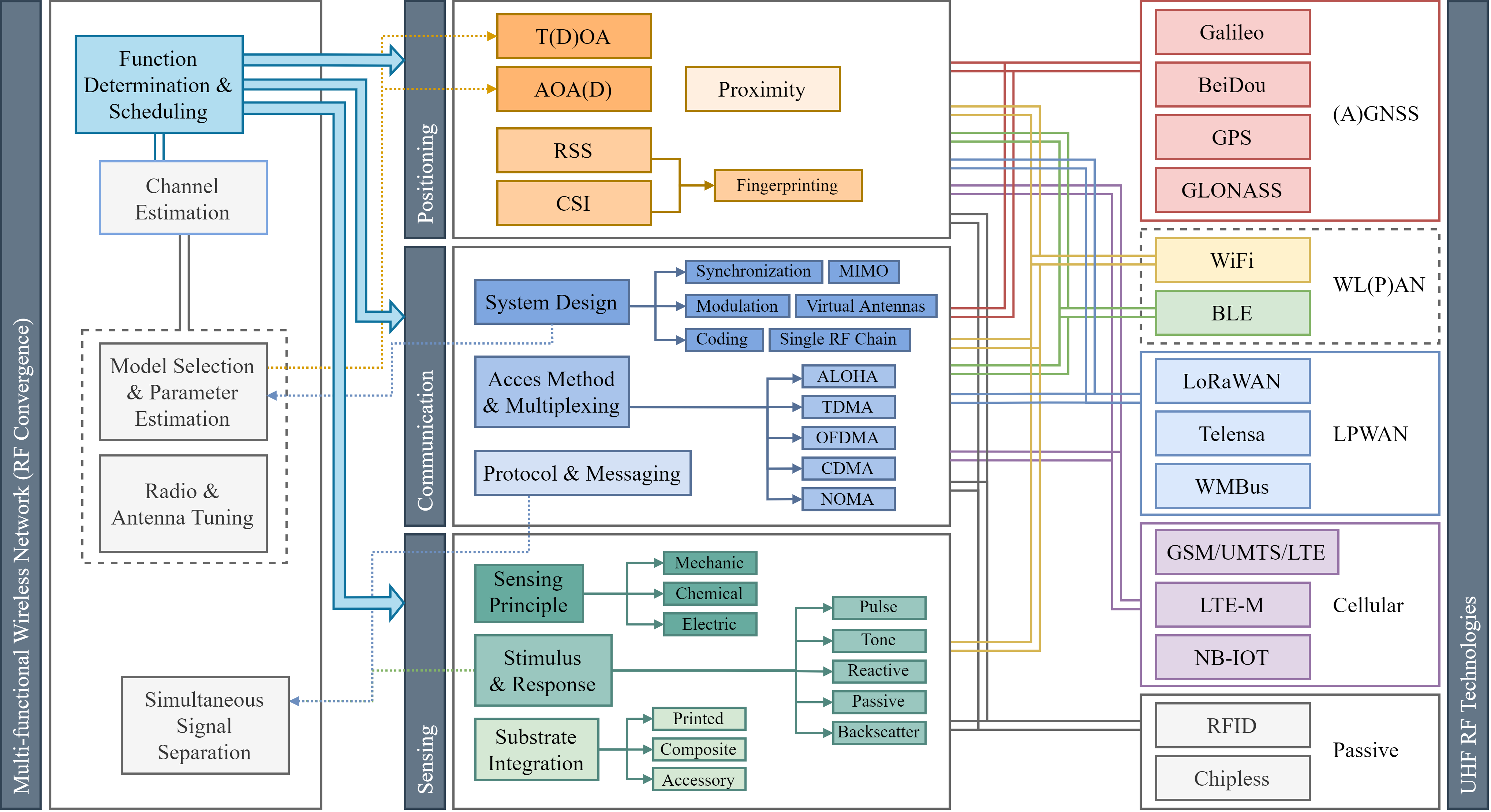} 
    \caption{Architecture for a multi-functional wireless network in UHF band. Within RFC concept, communication, positioning and sensing mutually depend on each other, and their execution may be synchronous or asynchronous. The joint functions are built on the protocol and metadata obtained by the radio.}
    \label{fig:fig1}
\end{figure*}

%%%%%%%%%%%%%%%%%%%%%%%%%%%%%%%%%%%%%%%%%%%%%%%%%%%%%%%%%%%%

%%
% Multifunctional Radio Systems: Communication, Positioning, and Sensing
%% 
\section{Multi-functional Radio Systems: Communication, Positioning, and Sensing}
\label{sec:multifuncradiosys}

% naming 
Depending on the target application, the RFC concept or multi-functional wireless networks is found with different terms in the literature: the most common are joint communication and sensing (JCAS) and joint communication and positioning (JCAP). However, all of them embrace a common background in which communication, positioning and sensing functionalities are executed by a single radio system. 
%intro to non-simultaneous vs simultaneous use
Yet, in the existing literature, it is difficult to find a specific differentiation between non-simultaneous and simultaneous use of radio functionalities. The latter being a more challenging approach in which multiple functions are executed simultaneously by giving each unit of RF energy spent a dual simultaneous use. Considering that most of the current approaches today are non-simultaneous, there are still many challenges and opportunities to explore the simultaneous functionality with these technologies. The simultaneous use of the network multiple functions may enable even further optimization, considering that the same energy derives into more than one relevant source of information for the application. Obtaining this simultaneous RF function is subject to numerous challenges, as discussed in Section~\ref{sec:challenges}.

%go into details of the building blocks.
RFC radio systems need to be orchestrated to perform any of the multiple functionalities, or even all of them together, when achieving full simultaneity. Fig.~\ref{fig:fig1} depicts such functionalities with the aim to illustrate the interrelation between different subsystems and the needed conceptual blocks. 
%\pt{Aqui es comença a pararlar de simultaneity però el concepte no s'ha introduit. Potser caldria moure aqui el concepte de simultaneitat. En concret: que vol dir simultenaitat i quins tipus hi ha? per a què es pot utilitzar la simultenaitat: comunicacions, posicionament i sensat}.
Primary and alternate functionalities may rely on the transported data in case of different network protocols execution, or data and metadata in case sensing functionalities are performed. The radio metadata (e.g., signal strength, phase, etc.) can be exploited by the application layer in order to derive secondary functions, usually leveraging signal models and channel estimated conditions and becoming a new scenario for AI-based approaches. This function can either be executed in the device or at the edge/cloud infrastructure, leveraging functional splitting architectures. The radio functionalities can be complemented with dedicated radio and antenna (de)tuning that may trigger or support the alternate function. When simultaneous functions are executed, information needs to be discriminated. Such discrimination may range from the separate processing of data and metadata, to the signal separation during the baseband processing steps. 

Today, simultaneity is not part of UHF radio systems. Hence, turning a radio system into a multi-functional one will require certain functional upgrades. In most cases, the transformation will not require any hardware upgrades, but only to include the appropriate processing logic, either embedded or at the edge/cloud.
Functions such as positioning and communication will be exploited in most of the UHF systems and, indeed, several standards provide some sort of these dual functions, mostly non-simultaneous (Section~\ref{subsec:apptrack}). Sensing in the UHF bands is still an emerging topic and has been approached by the Wi-Fi and RFID community (Section~\ref{subsec:apptrack} and \ref{subsec:embodied}). GNSS functionalities are being executed by cellular radios (e.g., Nordic NRF9160), and its simultaneous operation depends on the proper scheduling and duty-cycling of the functionalities as performed by a controller (Section~\ref{subsec:RAN}). 

%\gb{aixo ultim es la justificacio pq la seguent seccio es devideix en 3 subseccions? potser queda amagada?}

\begin{figure*}[]
    \centering
    \includegraphics[width=\linewidth]{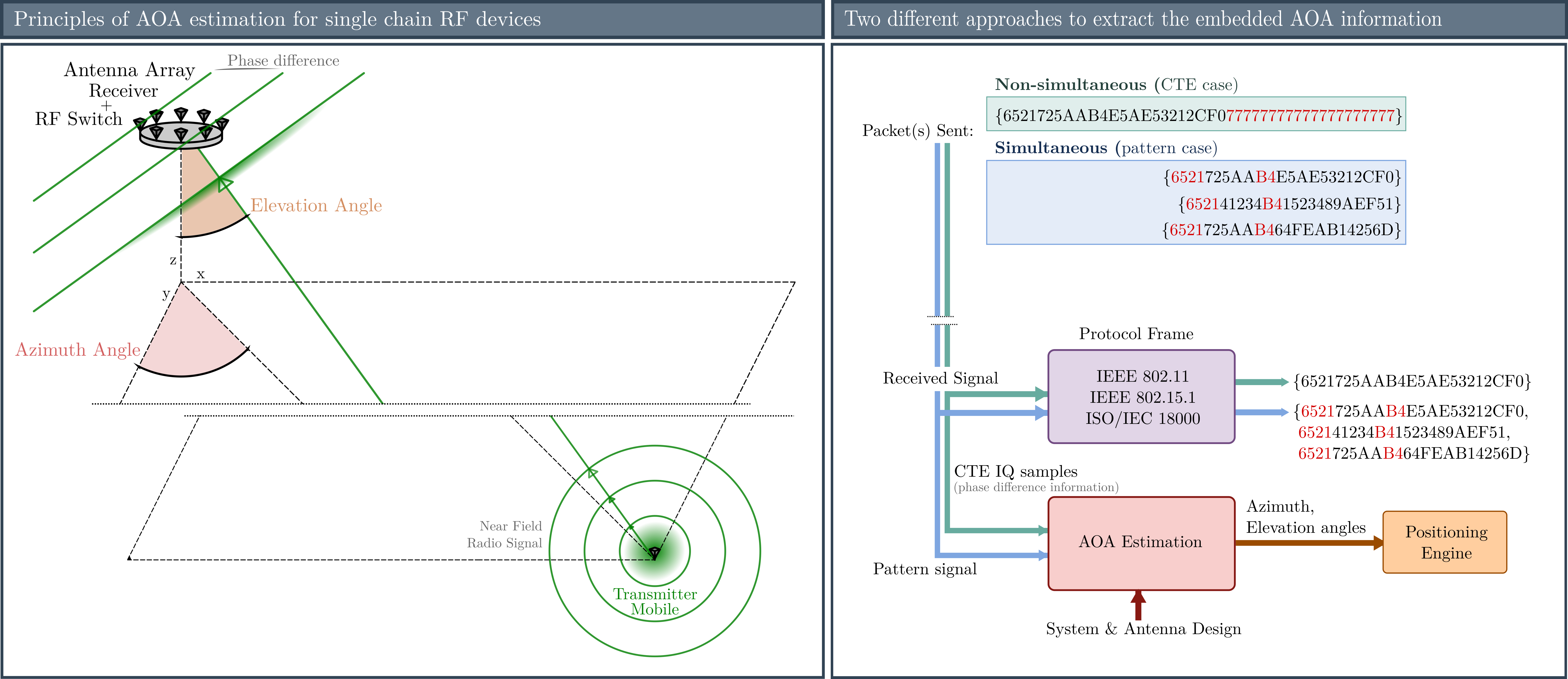} 
    \caption{Direction finding or AOA-based positioning using constant tone extensions (non-simultaneous) or recognizable patterns (simultaneous communication) for single RF chain devices. The incident signal arrives at each antenna at slightly different times, resulting in time and phase-delayed versions of the original that are processed in consideration of the antenna design to extract the corresponding embedded AOA information.}
    \label{fig:non-vs-sim}
\end{figure*}

%%%%%%%%%%%%%%%%%%%%%%%%%%%%%%%%%%%%%%%%%%%%%%%%%%%%%%%%%%%%

%%
% Applications of RF convergence in UHF band
%%
\section{Applications of RF convergence in UHF band}
\label{sec:applications}
This section presents the most relevant applications of RFC in the UHF band.
RFC applicability in the band may become interesting in scenarios in which the technology is part of a large and complex infrastructure, thus its replacement or extension may involve large economic investments. Particular use cases include industrial network deployments, critical infrastructure systems, and even building infrastructures. Other approaches may involve the embodiment of communication systems into materials and composites, providing multiple functionalities with simplified systems.

% Ubiquitous localization and sensing
\subsection{Ubiquitous localization and sensing}
\label{subsec:apptrack}

% Xavi, havies escrit aixo, no ho borro:
% The most common use of RF beyond communication is presence detection, localization, and tracking. RFID, Wi-Fi, and BLE are widely used technologies for that purpose in industrial settings, while LTE and LPWANs are introducing these features for large area coverage. Key to this possibility are the user equipment or terminals, mostly cell phones or small low-power tags widely adopted in the industry. Accuracy and area coverage are the main challenges, these are being addressed by the introduction of direction finding techniques and advanced modelling and pattern recognition via AI.  
%% figure aoa localization
%% buscar coses de wifi per localitzacio, em sona que tambe han estandardizat el direction finding

% Guillem:
%(bit error rate, channel capacity, data rate, etc.)  (accuracy, coverage, etc.)
The most common use of RF signals beyond communication is asset location. Verticals such as industry, healthcare, retail, agriculture, and cities need spectrum and energy-efficient (i.e., simultaneous) JCAP systems tailored to address their specific requirements. 
%Beneath each vertical lay many services, but most require uniquely identifying the asset, accurately knowing its position, and receiving its data; trading off latency tolerance and low data rates for low power, cost, and complexity. 6G is based on a unified signal to simultaneously optimize communication performance and positioning, but is clearly not the cost-effective solution for said verticals. 
Two current, but rather limited, simultaneous JCAP systems in UHF band are GNSS and RFID. In GNSS, the navigation signal is only used for outdoor positioning, while it also transmits reference positions. RFID integrates remote sensing, data transfer, detection, and proximity location to some extent (Section~\ref{subsec:embodied}). 
In contrast, other UHF technologies (e.g., Wi-Fi, BLE) that are already standardized and widely adopted by the industry and, thus, have a strong potential, but still have very immature simultaneous JCAP features.

% already include a set of mechanisms that are more adequate for simultaneous JCAP functionalities. However, the status  for such technologies is still unplanned/immature. 
% \pt{No s'enten: Instead, the already adopted and standardized technologies in UHF band include a set of more adequate but immature (in the simultaneous JCAS sense) candidates.} %, as such a simultaneous system does not yet exist.

%Unfortunately, such a multifunctional network system does not yet exist in the UHF band.  %The planned 6G is being efficiently designed from scratch to optimize communication performance and positioning, but there is no such system based on a unified signal structure that can simultaneously provide both services in the UHF band.
% aqui pq quedi a la 4 pagina

%On the other hand, (not so energy-efficient)
Parallel systems and extensions of communication systems are current alternatives to the lack of simultaneous JCAP systems. Examples of the former are single devices limited to outdoor positioning in open areas that combine GNSS with other RF communication technologies (e.g., LTE in mobile phones, and LoRa in remote LPWAN nodes). Examples of extensions are positioning methods exploiting signal measurements from different spatial sources, such as received signal strength (RSS), time on air (TOA), channel state information (CSI), angle of arrival (AOA), and angle of departure (AOD) of WLAN and LPWAN RF technologies. 
RSS or CSI-based fingerprinting offer simultaneous JCAP with limited accuracy, as the communication system is not specifically designed for positioning purposes~\cite{aranda2022performance}. 
TOA-based approaches require unfeasible large signal bandwidths in the UHF band for acceptable positioning performance. For example, to achieve a resolution of 30~cm, a bandwidth larger than 1~GHz is required~\cite{yang2015wifi}. %, 30~cm range resolution needs 1~GHz at least.
%, as well as precise time synchronization between transmitters and receivers.  %Thus, technologies like UWB in higher frequency bands are more suitable. 
%with 10 MHz bandwidth as well as sampling rate can only measure the time duration up to 1 × 10–7 s resolution. Therefore, the maximal error in distance is up to 3 × 10^8 × 10^–7 = 30 m
However, better positioning performance can be obtained with a proper redesign of the protocol signalling and waveform. A clear example of such approach outside the band are the new 5G features introducing Positioning Reference Signal (PRS) and the Sounding Reference Signal (SRS) for positioning in the downlink and uplink respectively. PRS and SRS use specific patterns for the receiver to determine the AoA of the signal. Similarly, BLE 5.1 added the Constant Tone Extension (CTE) to the BLE frame. This extension does not carry any data~\cite{ble1}, but allows the receiver to consecutively sample the symbols by switching between multiple antennas to determine the AoA information from the RF phase differences, thus mimicking multiple-input and multiple-output (MIMO) technology (Fig.~\ref{fig:non-vs-sim})~\cite{pau2021bluetooth}. Although being close to the strictly simultaneous JCAP with indoor sub-meter accuracy, there is still RF energy and channel occupancy intended for either communication or positioning, but not both simultaneously. 

Regarding simultaneous sensing capabilities, CSI analysis of Wi-Fi signals is currently being explored to sense, recognize, and detect environment variations with applications in imaging, human identification and gesture recognition, among others~\cite{he2020wifi}. Fine-grained characterization of the environment through CSI requires spatial diversity (e.g., MIMO) in conjunction with sub-carrier granularity (e.g., OFDM), thus Wi-Fi has emerged as a low-cost alternative solution to RF sensing in higher UHF frequency bands.
%Therefore, the widely available Wi-Fi emerges as a cost-effective alternative solution to RF imaging and indoor positioning with no hardware modification or redeployment required.
%human identification, gesture recognition, fall detection, and indoor positioning
%Wi-Fi imaging, vital sign monitoring, human identification, gesture recognition, gait recognition, daily activity recognition fall detection, human detection, and indoor positioning

% Embodied sernsors
\subsection{Embodied sensors}
\begin{figure}[!t]
    \centering
    \includegraphics[width=\linewidth]{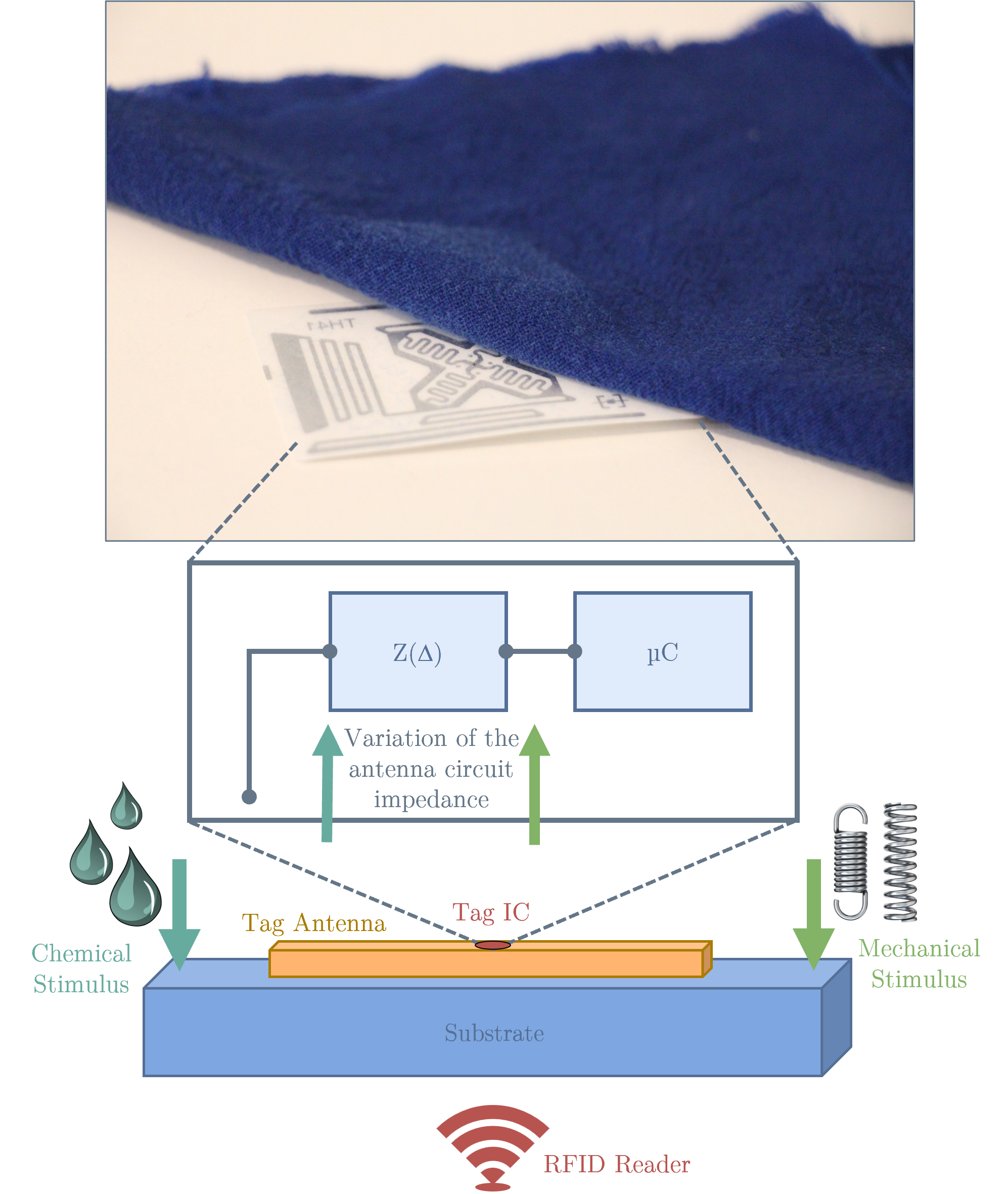} 
    \caption{Example RFID tag integrated into cotton fabric and used as moisture sensor~\cite{leja22}. The variation of the substrate condition (e.g., wet vs dry) causes a change in the antenna impedance that is materialized in a modification of the signal strength and phase perceived by the receiver. Other related behavior can be observed with mechanical stimulus to the substrate (e.g., elongation).}
    \label{fig:fig2}
\end{figure}

An emerging area of research with interesting applicability in the construction, recycling, consumer, and manufacturing industries is the embodiment of RF enabled sensors into parts, objects, or materials. For example, passive RFID tags can be embodied in concrete or attached to fabrics to enable non-intrusive, battery-less sensing (e.g., see Fig.~\ref{fig:fig2}).
Specifically, RFID communication is based on back-scattering, where the reader device initiates the communication, also providing a continuous RF wave, which is used by the tags to reflect the answer by means of a specific signal modulation. The fact that the reader device is able to compare the transmitted signal with the received one, allows an RFID system to use the same RF interface for communications (asset identification), and sensing through signal analysis. RF magnitudes like RSS, RF phase, or frequency shifting, can precisely be measured by commercial off-the-shelf equipment at no additional cost, enabling the sensing capacity of RFID~\cite{5508333}. 
Different applications can benefit from this simultaneous multi-functional use of this technology. For instance, UHF RFID has long been proposed as a position and movement sensor. This technology enables fine-grained localization since the reader can easily compare the transmitted RF phase with the received one thanks to the back-scattering nature of the communication. Coarse-grained localization and movement detection can also be achieved through RSS-distance modelling analysis. Moreover, an RFID reader can make use of a virtually unlimited number of antennas on a single RF front-end through time multiplexing, allowing spatial diversity on the signal reception.

Person-object interaction detection can also be achieved with similar principles than localization, also combining the effect of the human body in proximity to the tags, which may affect the performance of the antenna. Spatially distributed tags over objects or any surfaces allows a better understanding of the type of interaction (e.g., RFID tags antennas are short-circuited when in contact with human body).

RFID is also being used as a sensing mechanism for physical magnitudes like humidity, pressure or temperature, through antenna-based sensing. The sensing principle in this case relies on the changes in the antenna substrate, which in turn modifies the antenna-chip impedance matching, resulting in changes in the back-scattered signal. These changes can be tracked in time if the physical magnitude of interest can be isolated (as far as the reader keeps interrogating the tags), or by using differential analysis with isolated reference tags.

RFID tags, and specially chipless RFID, have the potential to become a truly ubiquitous sensing mechanism within the IoT ecosystem. On one hand, a single reader device can communicate with hundreds of tags per second, allowing this technology to scale at low cost. On the other hand, its virtually unlimited lifetime and low maintenance requirement, thanks to their battery-less nature and ultra low-cost (being similar to a barcode in the case of chipless RFID), puts this technology in a privileged position in the simultaneous multi-functional RF networks landscape.
\label{subsec:embodied}

% Simplified multi-RAN systems
\subsection{Simplified multi-RAN systems}
\label{subsec:RAN}

The development of systems able to operate different radio access technologies (RATs) simultaneously is a key cornerstone to achieve real multi-functional networks.
In that sense, the problem of simultaneous operation of more than a single RAT has been extensively addressed in the literature from different approaches, which can be classified into simultaneous RATs using the same frequency bands or using different frequency bands.
The former is aimed to optimize the performance of the coexisting technologies either by tuning the parameters of each technology, such as back-off periods, duty cycles, etc; or by exploiting cognitive ability of devices, sensing the environment and opportunistically accessing the medium, commonly known as cognitive networks.
As for the operation of RATs at different bands, research has focused on traffic offloading mechanisms, particularly in high traffic locations between technologies such as LTE-A and Wi-Fi.
However, the research in the field of real multi-functional networks, able to operate multiple RATs using a single RF front-end is still in its infancy, either in the same band or in different bands.

Currently, the resampling of the received signal followed by post-processing stages comes out as a possible solution to support the coexistence of several RATs at the same frequency band using a single RF front-end.
For example, in \cite{Mohamed2020}, the authors propose a detection method and a resampling technique to detect, receive and correctly process Wi-Fi, ZigBee and LTE operating in the 2.4~GHz band.
From a different approach, and exploiting the SDR potential, a virtualization layer has been proposed to allow the execution of multiple air interfaces simultaneously on a single shared RF front-end~\cite{Kist2018}.
Nevertheless, there is still room for improvement, mainly regarding the exploitation of the differences in structure of the protocols.
For instance, benefiting from idle or inactive RRC states of one technology to fit other technologies' transmissions or sensing/receiving periods, such as NB-IoT and GNSS, or even optimizing the energy use of the system in standards that require both positioning and communication, such as the IoT-NTN as proposed by the 3GPP Release 17\cite{3gpp}.
Also, duty cycling opens up opportunities to exploit such differences.

The challenge of using a single physical antenna in different bands has also been addressed as an enabler for multi-functional networks, with the design of antennas able to operate at distant bands.
A good example of this can be found in \cite{Adam2016}, where a single physical antenna aperture, consisting of more than one spatially interleaved antenna
arrays, can operate simultaneously and independently at bands such as Ka and Ku.
In a similar vein, Ignion~\cite{ignion} has launched a wide range of products exploiting the concept of \textit{virtual antenna}, multi-band, multipurpose and ready-to-use chip antennas developed for the IoT market.

%%%%%%%%%%%%%%%%%%%%%%%%%%%%%%%%%%%%%%%%%%%%%%%%%%%%%%%%%%%%

%%
% Challenges and Opportunities
%%
\section{Challenges and Opportunities}
\label{sec:challenges}

\begin{table*}[t]
\centering
\begin{tabular}{@{}lll@{}}
\cmidrule(r){1-2}
\multicolumn{1}{c}{\textbf{Non-Simultaneous}}     & \multicolumn{1}{c}{\textbf{Simultaneous}}      &                          \\ \cmidrule(r){1-2}
-- Coordination/Scheduling                           & -- Simultaneous Reception (Virtual Antennas)      &                          \\
-- Antenna switching                                 & -- Multi-RAN baseband processing                  & Multiple Functionalities \\

-- Reconfigurable RAN                                & -- Programmable Radios                             & on a Single Radio        \\
-- APIs and system integration                       & -- Interference Cancellation (NOMA)                &                          \\
-- Standardization                                   & --  Standardization                                &                          \\ \cmidrule(r){1-2}
-- Signal propagation modelling & -- Pattern recognition in protocol headers        &                          \\
-- Mapping of signal to network function       & -- Dealing with Security (e.g., Encryption)       & Exploiting Propagation \\
-- Pattern recognition methods                       & -- Edge processing for function separation        & and Protocol Features    \\
-- Protocol extensions                               & -- Dealing with physical layer randomization (Interleaving)                                               &                          \\
-- Overhead minimization (energy, duty cycle) &      
                                               &                          \\ \cmidrule(r){1-2}
-- Modelling electrical influence of substrate & -- Influence of substrate to the primary function &
\\
-- Protocol ext. for alternate function trigger      & -- Modelling and pattern recognition methods      &                          \\
-- Substrate influence on the primary function & -- Hardware integration into materials            & Physical Influence     \\
-- Identifying physical causes to be modelled        & -- Protocol extensions for simultaneous functions &                          \\
-- RF harvesting and battery-less systems            & -- RF harvesting and battery-less systems         &                          \\ \cmidrule(r){1-2}
\end{tabular}
\caption{Research challenges and opportunities for the development of RFC on existing radio systems.}
\label{tab:my-table}
\end{table*}

The development of RFC reusing an existing technological base is per se a challenge that embraces the design and implementation of these coexisting functionality on already designed technologies and protocols. Yet, the reuse and extension of existing network technologies brings important opportunities for a more complete digitization, while minimizing the need for new equipment. In what follows, we aim to identify the main challenges and outline research opportunities to contribute to the development of that trend. These findings are also summarized in Table \ref{tab:my-table}.

% Asset location using RF
\subsection{Asset location using RF}
As presented in Section~\ref{subsec:apptrack}, JCAP systems have already been implemented and even standardized. The evolution of network protocols towards simultaneous functionalities are key for a more integrated communication and positioning. Position estimation is based in the analysis of the signal characteristics, either by determining the distance from the signal strength or by determining the direction of the signal through exploiting models and filters that build on the phase variation. The latter is facilitated in single-input and single-output (SISO) or single RF chain systems by constant tones embedded in the frame, in order to simplify the determination of the phase variation as received from an antenna array.  
Simultaneous communication and position determination may require the identification of patterns in the protocol headers or payload (Fig.~\ref{fig:non-vs-sim}), minimizing the overhead in the frame due to padding or constant tones. Modelling and pattern matching research efforts may enable the extraction of unique signatures from frames and these used for the determination of the signal direction. Finally, Artificial Intelligence (AI) methods can play a relevant role in the identification of these patterns. 
Another challenge to overcome is the effect of encryption of frames in the identifiable patterns. Security architectures and protocols need to be extended for that purpose, even proposing multi-layer security credentials that can ensure the proper confidentiality while enabling simultaneous network functions. A similar effect may be achieved by the interleaving function in some physical layers. Addressing this randomization is a research challenge that may lead to more intelligent de-interleavers.
Finally, other directions involve the further exploitation of MIMO features of current radio systems to exploit antenna diversity for direction finding techniques.

% \xv{check.. no se si podem fer AoA amb antennes MIMO.}
% \gb{LORA AOA SIMO SDR: \url{https://ieeexplore.ieee.org/document/9269490}.}
% - \gb{distinguish between MIMO and SISO systems}
% - \gb{specific hardware (increased cost)}

% Physical parameter sensing with RF
\subsection{Physical parameter sensing with RF}
Sensing physical phenomena with radios is an emerging topic, especially for active radios. The sensing principle is still an open field for research in which the impact of different substrates and conditions (i.e., mechanical, chemical and electric) need to be characterized. %\pt{Costa llegir: While the effects in the electric impedance of the antenna circuit are barely explored, in certain areas, such as RFID, these are completely unexplored for active radios.} 
Beyond the variations of the antenna impedance, the effects of the physical phenomena (e.g., pressure, temperature, elongation, etc.) into the radio circuits have only been studied from a signal propagation perspective, and mostly in passive radios, and not analyzing the variation with respect to the communication conditions. Observing effects in the frequency (de)tuning, signal strength and phase, among others, may lead to novel sensing opportunities. More generally, the relations between the Micro-Electro-Mechanical (MEMs) and sensor design areas need to be further investigated to leverage learnings already in place on existing sensing technologies.  
There are also challenges that need to be overcome in the effect of physical phenomena into the communication technology (e.g., assessing its impact into the protocol structure at different layers). For example, for backscattering-based systems, the duration of the system interrogation may have significant impact in the sensing capability. In contrast, for active systems the impact of the physical phenomena may have incidence in the bit error rate and, thus, require more robust redundancy measures (e.g., coding). Yet, the compliance with regulations, as well as with the protocol or standard, may limit its applicability. Other unexplored areas may embrace sensing as part of the RF energy harvesting as alternate dual use. 

% 0.6 pagina
%- Effects of substrate on antenna behaviour
%    - modelling vs pattern matching
%    - Signal strength
%    - Phase
%    - Detunning
%    - Other
%        -SNR, etc..
%- Effects of mechanical aspects on antenna behaviour
%- Effects of environment on propagation
%   - e.g. cross material propagation
%- Active vs Passive systems
%- BAP tags
%- Chip vs. chipless
%- In passive systems, impact of protocol 
%   - duration of interrogation

% describe here how physical phenomena, e.g contact to different materials, environmental conditions have an effect into the antenna matching, signal propagation, etc.. and therefore these can be observed in the RF signal in forms of phase shift, received power, snr, etc...

% Multifunctional RANs
\subsection{Multi-functional RANs}
Using a single radio for multiple purposes becomes a requirement as systems become more integrated. For example, the 3GPP IoT-NTN Group is assuming GNSS connectivity at each user equipment (UE) in order to properly interact with non-terrestrial base stations. Using a single radio for that purpose (e.g., NB-IoT and GNSS) is possible as long as the proper functionality orchestration and antenna are in place. The challenges for multi-functional RANs using already established technologies are in the control plane, defining the proper policies for radio functionality coordination and scheduling, duty cycling and antenna switching when needed. Existing radios can benefit from virtual antenna designs, that can offer multiple simultaneous front-ends. More challenging is the orchestration of RANs for simultaneous operation, although this is possible if the stacks are fully softwarized (e.g., IEEE 802.15.1 and 802.15.4). Here, the main challenges are protocol signalling coordination and timing. Moreover, as some radios are already equipped with programmable components, such as DSPs or FPGAs, the exploitation of such hardware can enable alternate network functions. Exploiting a radio for multiple RANs may also require the use of differentiated antennas. This may trigger an evolution of antennas towards programmability and virtualization. At the physical and MAC layer, novel non-orthogonal modulations (e.g., NOMA) may open opportunities for simultaneous sensing and communication, especially when considering the different natures of sensing (slow changing) and communication (data intensive). In that regard, the feasibility of NOMA for this purpose on existing 4G and 5G infrastructures needs to be explored and validated. 

%%%%%%%%%%%%%%%%%%%%%%%%%%%%%%%%%%%%%%%%%%%%%%%%%%%%%%%%%%%%
%%
% Conclusions
%%
\section{Conclusions}
\label{sec:conclusions}
In this article, we have analyzed the challenges and opportunities for multi-functional radio technologies building on well-established communication technologies in the UHF bands for positioning, sensing and simultaneous communication applications. The article has presented the common use cases, existing solutions and directions for improved and simultaneous network functions.
Regarding positioning, we have found that non-simultaneous positioning and communication is already supported by most of existing radio technologies. However, simultaneous positioning and communication still needs to address important protocol-related and security challenges.
On the sensing side, RF sensing is still in its early days. While RFID and Wi-Fi have shown interesting approaches, with potential impact into both human and industrial processes, the concept still needs to be further developed in the active radio fields and linked to relevant research areas, such as MEMs. 
Other interesting research and development opportunities focus on using a single radio to access different RANs. Challenges such as handling the simultaneous reception, separation of signal or even studying the suitability of NOMA for that purpose are still to be addressed. 

In a more general sense, we aim to stress the opportunities and value that can be brought by extending existing wireless infrastructures with alternate functionalities, especially in infrastructures in which UHF technologies are already in place. The research community should look not only to the new opportunities brought by new technological advances and trends, but also to build and develop upon existing technologies, pushing the limits and overcoming barriers that were established decades ago.  
\bibliographystyle{IEEEtran}
\bibliography{references}

\begin{IEEEbiographynophoto}{Xavier Vilajosana}
(M'09, SM'15) received his B.Sc. and M.Sc in Computer Science from Universitat Politècnica de Catalunya (UPC) and his Ph.D. in Computer Science from the Universitat Oberta de Catalunya (UOC). He has been a researcher at Orange Labs, HP and UC Berkeley. He is now Professor at UOC. 
\end{IEEEbiographynophoto}

\begin{IEEEbiographynophoto}{Guillem Boquet}
is a Researcher (2020) at Wireless Networks (WiNe) group at Universitat Oberta de Catalunya (UOC) and Assistant Professor (2014) at Universitat Aut\`onoma de Barcelona (UAB). He holds an M.Sc. (2014) and PhD. (2021) in Telecommunications Engineering from UAB.
\end{IEEEbiographynophoto}

\begin{IEEEbiographynophoto}{Joan Meli\`a-Segu\'i}
received his B.Sc. and M.Sc. in Telecommunications Engineering from Universitat Politècnica de Catalunya and his Ph.D. from the Universitat Oberta de Catalunya (UOC). He has been researcher at the Universitat Pompeu Fabra and visiting researcher at the Palo Alto Research Centre (Xerox PARC). He is currently Associate Professor at the Faculty of Comp. Science, Multimedia and Engineering at UOC.
\end{IEEEbiographynophoto}

\begin{IEEEbiographynophoto}{Pere Tuset-Peiró}
(M'12, SM'18) is Associate Professor at the Universitat Oberta de Catalunya (UOC) and Senior Researcher at the Wireless Networks (WiNe) group. He received his M.Sc. in Telecommunications Engineering from Universitat Politècnica de Catalunya (UPC), and his Ph.D. in Network and Information Technologies from Universitat Oberta de Catalunya (UOC).
\end{IEEEbiographynophoto}

\begin{IEEEbiographynophoto} {Borja Martinez}
received his B.Sc. in physics %the M.Sc. degree in microelectronics 
and the Ph.D. in informatics from the Universidad Aut\`onoma de Barcelona (UAB), Spain, 
where he was assistant professor from 2005 to 2015. 
%combining this activity with applied research in the private sector. 
He is currently a research fellow at the IN3-UOC. 
His research interests include low-power wireless technologies and energy management policies. % and algorithms. 
\end{IEEEbiographynophoto}

\begin{IEEEbiographynophoto}{Ferran Adelantado} 
(M'08, SM'19) is Associate Professor at the Universitat Oberta de Catalunya (UOC) and Senior Researcher at the Wireless Networks (WiNe) group. He holds a M.Sc. degree in Telecommunications Engineering (2001) and a PhD (2007) from the Universitat Politècnica de Catalunya (UPC).
\end{IEEEbiographynophoto}

\end{document}